\catcode`\@=11
\def\gsim{\ifmmode{\mathrel{\mathpalette\@versim>}}
    \else{$\mathrel{\mathpalette\@versim>}$}\fi}
\def\lsim{\ifmmode{\mathrel{\mathpalette\@versim<}}
    \else{$\mathrel{\mathpalette\@versim<}$}\fi}
\def\@versim#1#2{\lower 2.9truept \vbox{\baselineskip 0pt \lineskip 
    0.5truept \ialign{$\m@th#1\hfil##\hfil$\crcr#2\crcr\sim\crcr}}}
\catcode`\@=12
\def\al2{\alpha^2}
\def\arctan{{\rm arctan}}
\def\arccosh{{\rm arccosh}}
\def\betah{\beta_{\rm h}}
\def\cmu{c_{\mu}}
\def\Cz{C_0}

\def\Cd{C_2}

\def\Cq{C_4}
\def\Cs{C_6}
\def\Cm{C_{\rm m}}
\def\csic{\xi_{\rm c}}
\def\d2{\delta^2}
\def\Dz{D_0}

\def\Dd{D_2}
\def\Dq{D_4}
\def\Ds{D_6}
\def\Dm{D_{\rm m}}
\def\eps{{\cal E}}
\def\Etil{\tilde\eps}
\def\fid{f_{\rm k}}

\def\ftil{\tilde f}
\def\Ftil{\tilde F}
\def\Ftilpm{\Ftil^{\pm}}
\def\Ftilpmis{\Ftilpm_{\rm i}}
\def\Ftilpman{\Ftilpm_{\rm a}}
\def\Ftilp{\Ftil^+}
\def\Ftilm{\Ftil^-}
\def\Ftilpis{\Ftilp_{\rm i}}
\def\Ftilmis{\Ftilm_{\rm i}}
\def\Ftilpan{\Ftilp_{\rm a}}
\def\Ftilman{\Ftilm_{\rm a}}
\def\fn{f_{\rm N}}
\def\fis{f_{\rm i}}
\def\fan{f_{\rm a}}
\def\Gtil{\tilde G}
\def\Gtilpm{\Gtil^{\pm}}
\def\Gtilp{\Gtil^+}
\def\Gtilm{\Gtil^-}
\def\gn{g_{\rm N}}
\def\gap{{\cal G}}
\def\k2{k^2}
\def\ln{\hbox{${\rm ln}\, $}}

\def\l2{l^2}
\def\mh{M_{\rm h}}

\def\muh{\mu _{\rm h}}
\def\parn{\par\noindent}
\def\psih{\Psi_{\rm h}}
\def\psin{\Psi_{\rm N}}
\def\psit{\Psi_{\rm T}}
\def\psii{\Psi_{\rm k}}
\def\psitil{\tilde\Psi}
\def\psitilm{\psitil_{\rm M}}
\def\psitilmu{\psitilm (\mu)}
\def\psitilh{\tilde\psih}
\def\psitilt{\tilde\psit}

\def\qi{Q_{\rm k}}
\def\Qtil{\tilde Q}
\def\ra{r_{\rm a}}
\def\rai{r_{\rm ak}}

\def\rc{r_{\rm c}}
\def\rch{r_{\rm h}}
\def\rhoh{\rho _{\rm h}}
\def\rhotil{\tilde\rho}
\def\roi{\rho _{\rm k}}
\def\roqi{\varrho _{\rm k}}

\def\rhon{\rho_{\rm N}}
\def\rap{{\cal R}}
\def\rapis{\rap_{\rm i}}
\def\rapan{\rap_{\rm a}}
\def\sa{s_{\rm a}}
\def\sad{\sa^2}
\def\sat{\sa^3}
\def\saq{\sa^4}
\def\sas{\sa^6}
\def\sac{s_{\rm ac}}
\def\Krad{K_{\rm rad}}
\def\KradH{\Krad^{\rm H}}
\def\KradBH{\Krad^{\rm BH}}
\def\Ktan{K_{\rm tan}}
\def\un{U_{\rm N}}
\def\vab{|b|}
\def\Vz{V_0}
\def\Vu{V_1}
\def\Vd{V_2}
\def\Vt{V_3}
\def\Vq{V_4}
\def\Vm{V_{\rm m}}
\def\vrad{v_{\rm r}}
\def\vtan{v_{\rm t}}
\def\sn{{\rm sn}}
\def\cn{{\rm cn}}
\def\dn{{\rm dn}}
\def\nc{{\rm nc}}
\def\tn{{\rm tn}}
\def\Xz{X_0}
\def\Xu{X_1}
\def\Xd{X_2}
\def\Xt{X_3}
\def\Xq{X_4}
\def\Xm{X_{\rm m}}

\documentstyle[11pt,aaspp4]{article}

\slugcomment{In press on Ap.J., Main Journal}

\lefthead{Ciotti L.}
\righthead{The Distribution Function of Two-Components Hernquist Models}

\begin{document}

\title{The Analytical Distribution Function\\
    of Anisotropic Hernquist+Hernquist Models}

\author{L. Ciotti}
\affil{Osservatorio Astronomico di Bologna, via Zamboni 33, 40126 Bologna, 
       Italy}



\begin{abstract}

The analytical phase-space distribution function (DF) of spherical 
self--consistent galaxy (or cluster) models, embedded in a dark matter halo, 
where both density distributions follow the Hernquist profile, with different 
total masses and core radii (hereafter called HH models), is presented. The 
concentration and the amount of the stellar and dark matter distributions are 
described by four parameters: the mass and core radius of the {\it reference} 
component, and two dimensionless parameters describing the mass and core radius
of the {\it halo} component. A variable amount of orbital anisotropy is allowed
in both components, following the widely used parameterization of 
Osipkov-Merritt. An important case is obtained for a null core radius of the 
halo, corresponding to the presence of a central black hole (BH). 

Before giving the explicit form for the DF, the necessary and sufficient 
conditions that the model parameters must satisfy in order to correspond to a 
{\it consistent} system (i.e., a system for which each physically distinct 
component has a positive DF), are analytically derived. In this context it is 
proved that globally isotropic HH models are consistent for any mass ratio and
core radii ratio, even in the case of a central BH. In this last case the 
analytical expression for a lower limit of the anisotropy radius of the host 
system as a function of the BH mass is given. These results are then compared 
with those obtained by direct inspection of the DF. In the particular case of 
global isotropy the stability of HH models is proved, and the explicit formula
for the differential energy distribution is derived. Finally, the stability of
radially anisotropic HH models is briefly discussed. 

The expression derived for the DF is useful for understanding the relations 
between anisotropy, density shape and external potential well in a consistent 
stellar system, and to produce initial conditions for N-body simulations of 
two-component galaxies or galaxy clusters.
\end{abstract}




%

\section{Introduction}


Recent ground based observations (\cite{msz95}), and with the Hubble Space 
Telescope show that the spatial luminosity distributions of elliptical galaxies
approach a power-law form $\rho(r)\propto r^{-\gamma}$ at small radii, with 
$0\leq\gamma\leq 2.5$ (\cite{cra93,jaf94,fer94,lau95,kor95,byu96}). These 
findings increase considerably the interest of theorists in the study of cuspy
models. Two important families of spherical dynamical models with a central 
divergent density that have been explored so far are the $R^{1/m}$ models and 
the so-called $\gamma$-models. The dynamical properties of models whose surface
brightness distribution follows the $R^{1/m}$-law, introduced by Sersic (1968)
as a natural generalization of the de Vaucouleurs law (\cite{dev48}), have been
extensively studied (\cite{cio91,cl96}). Particularly, their deprojected 
density increases toward the center as $r^{-(m-1)/m}$ for $m>1$; unfortunately
two major problems afflict these models: their deprojected density cannot be 
expressed analytically in terms of known functions, and no galaxies with 
$\gamma >1$ can be accurately modeled in their central regions. The family of 
the $\gamma$-models, in some way anticipated by Hernquist (\cite{her90}, 
hereafter H90), has been widely explored (\cite{deh93,car93,tre94}) and it 
represents a generalization of the well known Hernquist (H90) and Jaffe 
(\cite{jaf83}) density distributions. As shown by the previous authors, many of
the dynamical properties of the $\gamma$-models can be expressed analytically.
In particular the Hernquist model (hereafter H model) in projection well 
resembles the de Vaucouleurs law, and an exhaustive analytical investigation of
its properties is possible (H90).

It is now accepted that a fraction of the mass in galaxies and clusters of 
galaxies is made of a dark component, whose density distribution differs from 
that of the visible one. The shape of the dark matter distribution is not well
constrained by observations, but numerical simulations of dissipationless 
collapses seem to favor a peaked profile, consistent with the scale-free 
nature of the gravitational field (\cite{dc91,white96}, 
and references therein). 
From these considerations it follows that the obvious generalization of the 
one-component spherical models (the dynamicists zero-th order approximation of
real galaxies) is not only in the direction of the actively developed modeling
of axisymmetric and triaxial systems [see, e.g., de Zeeuw (1996) for a recent 
review] but also in the study and construction of two-component analytical 
models, a field far less developed. From this point of view the zero-th order
approximation of realistic galaxies is the construction of analytical 
spherically symmetric {\it two-component} galaxy models. 
 
When studying a dynamical model (single or multi-component) the fact that the 
Jeans equations have a physically acceptable solution is not a sufficient 
criterion for the validity of the model: the essential requirement to be met by
any acceptable dynamical model is the positivity of the DF of each physically 
distinct component. A model satisfying this minimal requirement (much weaker 
than the model stability) is called a {\it consistent} model. Two general 
strategies can be used to construct a consistent model or check whether a 
proposed model is consistent: the ``$f$ to $\rho$'' and the ``$\rho$ to $f$'' 
approaches (\cite{bt87}, Chap. 4, hereafter BT87). An example of the first 
approach is the extensive survey of two--component, spherical, self--consistent
galaxy models carried out by Bertin and co-workers (\cite{bss92}). They assume
for the stellar and dark matter components two distribution functions of the 
$f_{\infty}$ form (and so positive by choice) (\cite{bs84}). The main problem 
with this approach is that generally the spatial density is not expressible in
terms of known functions, and so only numerical investigations are feasible.

In the second approach the density distribution is given, and assumptions on 
the model internal dynamic are made, making the comparison with the data 
simpler. But the difficulties inherent in the operation of recovering the DF in
many cases prevent a simple consistency analysis. In particular, in order to 
recover the DF of spherical models with anisotropy two techniques have been 
developed from the original Eddington (1916) method for isotropic systems: the
Osipkov-Merritt technique (\cite{osi79,mer85}, hereafter OM), and the case 
discussed by Cuddeford and Louis (\cite{culo95}, and references therein). 
Examples of {\it numerical} application of the OM inversion to two-component 
spherical galaxies can be found in the literature (see, e.g., \cite{cp92}, 
hereafter CP92; \cite{czm95}). For axisymmetric systems recently a new 
inversion technique, less restrictive than the classical ones 
(\cite{lyb62,hun75,dej86}), has been found (\cite{hq93}). If one is just 
interested in the consistency of a stellar system the previous methods give 
"too much", i.e., give the DF. A simpler approach, at least for spherically 
symmetric multicomponent systems with OM anisotropy -- as the case discussed in
this paper -- is given by a method described by CP92, that requires 
information only on the radial density profiles of each component.

Despite all these efforts, a small number of one-component systems in which 
both the spatial density and the DF are analytically known is at our 
disposition, and in the more interesting case of two-component systems only the
very remarkable axisymmetric Binney-Evans model is known (\cite{bin81,eva93}).
It is therefore of particular interest the result here proved that also the DF
of HH models with OM anisotropy is completely expressible in an analytical 
way. 
This family of models is made by the superposition of a stellar and a dark 
matter distribution both following the Hernquist profile, with different total
masses and core radii. The concentration and the amount of the stellar and dark
matter distributions are described by four free parameters, and the orbital 
anisotropy is allowed in both components, following the OM prescription. A 
particularly interesting case is obtained for a null core radius of the "halo",
so mimicking a central BH. The study of HH models is also useful for many 
different reasons: to provide an analytical DF for a two-component cuspy system
for which the analytical solution of the Jeans equations is also available 
(\cite{clr96}); to investigate the r\^ole of anisotropy and mass distribution 
of each component in determining the positivity of their DF; to compute in an 
accurate and "easy" way the model line-profiles, to arrange initial conditions
for numerical simulations of two-component systems.

In Section 2 I briefly review the method presented in CP92, formulating it in a
way suitable for its application to the present problem. Then in Section 3 I
introduce the HH models, and use the previous method to discuss the limits 
imposed on their parameters by the positivity of the DF of the two components.
It is proved that globally isotropic HH models are consistent for any mass 
ratio and core radii ratio, even in the case of a central BH. In this last case
the analytical expression for a lower limit of the anisotropy radius of the 
host system as a function of the BH mass is given. In Section 4 I derive the DF
for HH models, and their differential energy distribution in the case of global
isotropy; some velocity sections of the DF are also shown. In the case of a 
dominant halo it is found that the DF of an HH component can be expressed only
through elementary functions. A particular case -- corresponding to a
Hernquist model with a central BH -- is extensively discussed. In Section 5 the
exact boundary of the region of consistency in the parameter space is obtained
using the DF, and the results are compared with those given in Section 3. In 
the same section the stability of globally isotropic HH models is proved, and a
discussion on the stability of the anisotropic case is given. Finally in 
Section 6 the main results are summarized.

\section{The Consistency of Multi-Component Systems}

More important than the construction of its spatial and projected velocity 
dispersion profile is checking whether a galaxy (or a galaxy cluster) model 
is consistent, i.e., is described by a DF everywhere non-negative. If a system 
is described as a sum of different density components $\roi$, then {\it each} 
$\fid$ must be non negative. This requirement leads us to introduce the 
concept of consistent multi-component decomposition of a system, as discussed 
in CP92, together with the main theorem used here. 
This theorem permits us to check whether
the DF of a multi-component spherical system where the orbital anisotropy of 
each component is described by the OM anisotropy is positive, {\it without} 
calculating it effectively. In the OM formulation the radially anisotropic case
is obtained as a consequence of assuming $f=f(Q)$ with: 
\begin{equation}
Q=\eps-{L^2\over 2\ra^2},
\end{equation}
where $\eps$ and $L$ are respectively the relative energy and the angular 
momentum modulus per unit mass, $f(Q)=0$ for $Q\leq 0$, and $\ra$ is the 
so-called {\it anisotropy radius}. With this assumption the models are 
characterized by radial anisotropy increasing with the galactic radius, and in 
the limits $\ra\to\infty $ the velocity dispersion tensor is globally 
isotropic. For a multi-component spherical system, the simple relation between
energy and angular momentum prescribed by equation (1) allows to express the DF
of the k--th component as:
\begin{equation}
\fid (\qi)={1\over\sqrt{8}\pi^2}{d\over d\qi}\int_0^{\qi} 
{d\roqi\over d\psit}{d\psit\over\sqrt{\qi-\psit}}=
{1\over\sqrt{8}\pi^2}\int _0^{\qi} 
{d^2\roqi\over d\psit^2}{d\psit\over\sqrt{\qi-\psit}},
\end{equation}
where 
\begin{equation}
\roqi =\roi\times \left (1+{r^2\over\rai^2}\right),
\end{equation}
$\psit (r)=\sum\psii (r)$ is the relative total potential, 
$\qi =\eps-L^2/2\rai^2$, and $0\leq\qi\leq\psit (0)$. The second equivalence 
in equation (2) holds for untruncated systems with a finite total mass (see, 
BT87, p.240), as the HH models here discussed. The original theorem, given in 
CP92 in terms of the model radius, is here formulated in terms of the relative
potential $\psii$ of the investigated component, since this formulation makes 
the treatment of HH models easier.
\medskip
\parn
{\bf Theorem }: Necessary condition for the non negativity of $\fid$ given in 
equation (2) is:
\begin{equation}{d\roqi\over d\psii}\geq 0,\quad 0\leq\psii\leq\psii(0).
\end{equation} 
If this necessary condition is satisfied, a {\it strong sufficient condition} 
(SSC) for the non negativity of $\fid$ is:
\begin{equation}
{d\over d\psii}\left[{d\roqi \over d\psii}
\left({d\psit\over d\psii}\right)^{-1}\sqrt {\psit}\right]\geq 0,
\quad 0\leq\psii\leq\psii(0).
\end{equation}
\parn
{\bf Proof}: See CP92.

Note that a {\it weak sufficient condition} (WSC) 
\begin{equation}
{d\over d\psii}\left[{d\roqi \over d\psii}
\left({d\psit\over d\psii}\right)^{-1}\right]\geq 0
\end{equation}
is obtained in a much easier way using the last expression in equation (2), 
requiring that ${d^2\roqi\over d\psit^2}\geq 0$. The WSC is obviously better 
suited than the SSC for analytical investigations, due to the absence of the 
weighting square root of the total potential. 

Some considerations follow looking at the previous conditions. The first is 
that the violation of the necessary condition [eq. (4)] is connected only with
the radial behavior of $\roi$ and the value of $\rai$, and so this condition is
valid {\it independently} of any other interacting component added to the 
model. Even when the necessary condition is satisfied, $\fid $ can be negative,
due to the radial behavior of the integrand in equation (2), which depends on 
the total potential, on the particular $\roi$, and on $\rai$: some permitted 
values of $\rai$ satisfying the necessary condition must be discarded. 
Naturally, the true critical anisotropy radius is always larger than or equal 
to that given by the necessary condition, and smaller than or equal to that 
given by SSC and WSC. The previous analysis has been performed obtaining 
analytical or numerical limits on $\ra$ for some widely used models: for 
example, in CP92 the King (1972), de Vaucouleurs (1948), and quasi-isothermal 
density distributions were discussed, and more recently the Jaffe (1983), 
Hernquist (H90) and Plummer (1911) distributions (\cite{clr96}). 

A complete analysis of the HH models is given in the next paragraph, deriving 
the analytical constraints on $\ra$ for 1) an H model, 2) an H model with a 
central BH, and finally 3) showing the consistency of the general HH model in 
the case of global isotropy.

\section {The HH Models}

In the study of the HH models, some simplification arises from the fact that 
both components are described by the same functional form. So, I do not 
distinguish between "stars" and "dark matter", but simply between a 
{\it reference} system and a {\it halo} system. The mass $M$ and the core 
radius $\rc$ of the reference system are the normalization constants, so that 
its density distribution is:
\begin{equation}
\rho (r)\equiv\rhon\rhotil (s)={\rhon \over s(1+s)^3},
\end{equation}
where $s=r/\rc$ and $\rhon=M/2\pi\rc^3$ (H90). The halo density is described by
another Hernquist distribution, of mass $\mh=\mu M$ and core radius 
$\rch=\beta\rc$:
\begin{equation}
\rhoh (r)={\rhon \mu\beta\over s(\beta+s)^3}.
\end{equation}
Then the HH density profiles are fully determined by fixing the four 
independent parameters $(M,\rc,\mu,\beta)$, with $0\leq\mu$ and $0\leq\beta$. 
Note that for $\mu=0$ the HH models reduce to the H model, and that for 
$\beta\leq 1$ the halo density is more concentrated than the reference 
component.

The fundamental ingredient in recovering the DF is the relative potential 
$\Psi$, that for the reference component is 
\begin{equation}
\Psi (r)\equiv\psin\psitil (s)={\psin\over 1+s},\quad 0\leq\psitil\leq 1,
\end{equation}
and for the halo
\begin{equation}
\psih (r)\equiv\psin\psitilh(s)={\psin \mu\over \beta+s},
\end{equation}
with $\psin=GM/\rc$ (H90). Note how for $\beta =0$ the halo potential is that
of a BH of mass $\mh$ placed at the center of the reference system. As it will
become clear in \S 4, a fundamental property of HH models is that their total 
potential can be expressed as a simple function of the reference potential, 
namely 
\begin{equation}
\psitilt=\psitil\times\left (1+{\mu\over 1+ b\psitil}\right),
\quad b\equiv\beta -1\geq -1,\footnote{The case $b=0$ (i.e. $\beta=1$) is 
discarded from now on, it being the case of an H model of total mass 
$M(1+\mu)$, fully discussed in H90.}
\end{equation}
where the interval $0\leq\psitil\leq 1$ is monotonically mapped onto 
$0\leq\psitilt\leq 1+\mu/\beta$.

\subsection{The Necessary and Sufficient Conditions for the H Model}

Here I apply first the necessary condition to the H model to determine the 
{\it critical} anisotropy radius such that a higher degree of radial OM 
anisotropy produces a negative DF, no matter what kind of halo density 
distribution is added. The first step is to express the modified density 
$\varrho$ [see eq. (3)] as a function of the relative potential $\psitil$, 
obtaining:
\begin{equation}
\tilde\varrho (\psitil)={\psitil^4\over 1-\psitil}
\left[1+{(1-\psitil)^2\over\sa^2\psitil^2}\right],
\end{equation}
where $\sa=\ra/\rc$ is the dimensionless anisotropy radius. 

As shown in Appendix A [eqs. (A1)-(A2)], the necessary condition can be treated
analytically, requiring
\begin{equation}
\sa\geq\sqrt{{(3\psitilm-2)\over (4-3\psitilm)}}{(1-\psitilm)\over\psitilm}
\simeq 0.128,
\end{equation}
where $\psitilm$ is the value of the potential for which the r.h.s. of equation
(A1) is maximum. In Fig. 1 the solid line at the bottom represents the derived
lower bound for the anisotropy radius. 
 
Moving to the study of the sufficient conditions, I apply first the SSC. The 
calculations can be performed analytically, and the result is:
\begin{equation}
\sa\geq\sqrt{
{3(5\psitilm-2)(1-\psitilm)^3\over 15\psitilm^2-39\psitilm+28}
}{1\over\psitilm}\simeq 0.25,
\end{equation}
[eqs. (A3)-(A5)], a limit obviously higher than that obtained from the 
necessary condition. The true limit on $\sa$ for the H model is within the 
two previous values, and in fact its value determined directly from the DF is 
$\simeq 0.202$ (see \S 4.2). In Fig. 1 this value is represented by the solid 
line in the middle. The result of the application of the WSC to the H model 
will be obtained as a limiting case of the more general analysis done in the 
next paragraph.

\subsection{Sufficient Conditions for the H+BH and Isotropic HH Models}

In order to proceed further with this analytical discussion, and to consider 
the more complicated case of the presence of the halo, we need to use the WSC 
rather than the SSC, due to the presence of the square root of the total 
potential in the SSC. I now prove two important results: 

\begin{enumerate}
\item in the case of a H+BH model (i.e., an H model with a central BH of mass 
$\mu M$) the WSC permits us to recover analytically a minimum value of the 
anisotropy radius as a function of the central BH mass; 

\item {\it globally isotropic} HH models can be consistently constructed for 
{\it any value} of $(\mu,\beta)$. Particularly this means that {\it a globally
isotropic H model can consistently host a BH of any mass at its center.}
\end{enumerate}

In the case of a central BH the WSC prescribes that:
\begin{equation}
\sa\geq\sqrt{-{3\psitilmu^4-10\psitilmu^3+6(2+\mu)\psitilmu^2-
6(1+\mu)\psitilmu+(1+\mu)\over 3\psitilmu^2-8\psitilmu+6+6\mu}}
{1\over\psitilmu},
\end{equation}
[eqs. (A6)-(A10)], and corresponds to the uppermost solid line in Fig. 1. As 
intuitive, an increase of the BH mass produces an increase of the critical 
anisotropy radius, i.e., a decrease in the maximum value allowed for the radial
anisotropy. Two interesting consequences can be obtained performing the 
asymptotic analysis of the previous equation, for $\mu\to\infty$ and 
$\mu\to 0$.

The asymptotic behavior of equation (15) for $\mu\to\infty$ is $\sa(\mu)\geq 
1/\sqrt{2}+O(1/\mu)$, and so just a slight reduction in anisotropy with respect
to the one-component model is required by the presence of a BH of any mass; and
the finite value of $\sa(\infty)$ implies that a globally isotropic model can 
host a central BH of any mass. The limiting case of the previous analysis for 
$\mu=0$ is the WSC applied to the H model, and completes the discussion given 
in \S 3.1. The limit for $\mu\to 0$ of equation (15) is
\begin{equation}
\sa\geq\sqrt{{(3\psitilm-1)(1-\psitilm)^3\over
3\psitilm^2-8\psitilm+6}}{1\over\psitilm}\simeq 0.31,
\end{equation}
and the comparison of this value with the limit on the anisotropy radius 
derived from the SSC ($\sa\geq 0.25$) is instructive: the difference is due to 
the weight factor of the square root of the potential, contained in the SSC and
absent in the WSC. The numerical evaluation of the SSC in the case of a BH is 
very easy, and the result is plotted in Fig. 1 (dot-dashed line).  

A second interesting case for which the WSC can be treated analytically is that
of a completely isotropic component of the HH model: in Appendix A it is shown
that its DF is positive for any choice of $(\mu,\beta)$. Incidentally, this 
gives another proof that it is always possible to couple consistently a BH of 
any mass with a globally isotropic H model, in accordance with the previous 
result. 

\placefigure{fig1}

\section{The DF of HH Models}

After the preliminary discussion we can now proceed to the explicit recovering
of the DF. Due to the fact that both density components of the HH model are 
described by the same functional form, it suffices to compute the DF for the 
reference component with generic $(\mu,\beta)$, and then also the DF for the 
halo component is easily recovered. As for the density and the potential, also
for $f$ it is useful to work with dimensionless functions. So, for the 
reference component $f=\fn\ftil(\mu,\beta;\Qtil)$ with $\fn=\rhon\psin^{-3/2}$,
and $0\leq\Qtil\equiv Q/\psin\leq 1+\mu/\beta$. For the halo component a 
similar functional form holds, where the two dimensional constants are now the
mass and the core radius of the halo, and the two dimensionless parameters are 
$\muh\equiv M/\mh=1/\mu$ and $\betah\equiv\rc/\rch=1/\beta$. The DF for the 
halo component is then obtained from that of the reference component changing 
the normalization constant $\fn$ to $f_{\rm Nh}$, substituting the 
dimensionless parameters $(\mu,\beta)$ with their inverses, and re-scaling the
parameter $Q$ to the halo central potential. 

The easiest way to compute the DF is to use the first of the identities in 
equation (2). For the evaluation of the integral one would be tempted to 
express $\varrho(\psit)$ eliminating the radial coordinate from the modified 
density and the total potential: this can be formally done, but the resulting 
expression for the radius involves a quadratic irrationality, that after 
insertion in equations (3) and (7) produces an intractable expression. Here I 
follow another approach: instead of eliminating the radius, the variable of 
integration is changed from the total potential to the potential of the 
reference component. This is equivalent to a remapping of the domain of 
definition of the DF, from the range of variation of $\psit$ to the range of 
variation of $\Psi$, and leads to introduce a new parameter $q$, defined from 
equation (11) as:
\begin{equation}
\Qtil=q\times\left (1+{\mu\over 1+bq}\right),\quad 0\leq q\leq 1.
\end{equation}
With this change of variable, the DF is given by equations (B1)-(B3), but as 
shown subsequently in Appendix B, this is again {\it not} the {\it natural} 
parameterization for $f$, that is finally obtained defining the variable:
\begin{equation}
\l2\equiv 1+bq.
\end{equation}
After normalizing to the dimensional scales of the reference component, its DF
can be formally written as:
\begin{equation}
f(Q)\equiv\fis(Q)+{\fan(Q)\over\sa^2}=
{\fn\over\sqrt{8}\pi^2}\left({d\Qtil\over dl}\right)^{-1}
{d\over dl}\left[\Ftilpmis (l)+{\Ftilpman (l)\over\sa^2}\right];
\quad l=l^{-1}(\Qtil),
\end{equation}
where the subscripts refer to the isotropic and anisotropic parts of the DF 
respectively, and
\begin{equation}
{d\Qtil\over dl}=2{l^4 +\mu\over b l^3}.
\end{equation}
The sign $\pm$ in equation (19) corresponds to the case $b>0$ and $-1\leq b <0$
respectively, and $\Ftilpmis (l)$ and $\Ftilpman (l)$ are given in Appendix B.
Note how with the followed procedure $f(Q)$ results from the elimination of 
the parameter $q$ between equations (17) and (18)-(19). A first (but 
algebraically cumbersome) check of the derived formula is obtained evaluating 
analytically its limit for $\mu=0$, and recovering the DF of the H model given
by H90. The lengthy proof is not given here. In the general case of $\mu >0$, 
a check is obtained by confrontation of the analytical DF with that derived by
direct numerical inversion of equation (2): the two families of curves are 
indistinguishable, with percentual errors smaller than $10^{-5}$. 

In Fig. 2 (upper panel) some DFs are shown in the case of global isotropy. 
Their main characteristic is that when the halo is more extended than the 
reference component (dotted lines) they are more peaked than the DF of the H 
model (solid line). On the contrary, for haloes more concentrated than the 
reference component (short-dashed lines), the DFs are flatter. A particular 
case is that corresponding to a central BH (long-dashed lines): the DFs are 
positive -- as discussed in \S 3.2 -- but {\it not} monotonically increasing 
near the model center. In the lower panel of Fig. 2 the DF for the same models
above are shown, but in this case the anisotropy radius is fixed to $\sa=1$. 
The same qualitative comments as in the isotropic case apply.

\placefigure{fig2}

An important feature of Fig. 2 is the radically different behavior of $f$ for
$q\to 1$: while for any finite core radius of the halo the DF of the reference
component diverges near the center, in the presence of a central BH the DF 
converges, i.e., the DF for the BH case cannot be obtained {\it directly} as a 
limit for $\beta\to 0$ of the DF with $\beta >0$. The discontinuity in the DF 
behavior at high energies is due to the coefficient of the function $\Xd$ in 
equation (B12): the limit for $\beta\to 0$ of the product between the 
coefficient and $\Xd$ does not vanish, on the contrary, for $\beta=0$ this term
is zero. The reason for this behavior is that the halo potential is uniformly
continuous on $s\in [0,\infty[$ for $\beta >0$, but for $\beta=0$ the uniform 
continuity is lost, and so the equivalence of the limit for $\beta\to 0$ before
and after integration is not guaranteed anymore.

A more direct way to look at this discontinuity is to perform an asymptotic 
expansion of the DF for $q\to 1$. A brute force (and highly error prone) 
approach would be the expansion of the expressions given in Appendix B for 
$q\to 1$, but a cleaner asymptotic expansion can be obtained instead expanding
directly equations (B1)-(B3): for $\beta >0$ the leading term is
\begin{equation}
\ftil(q)\sim{3\over 8\sqrt{2}\pi}{[1+(\beta-1)q]^2\over \mu+[1+(\beta-1)q]^2}
{\beta\over\sqrt{\mu+\beta^2}(1-q)^{5/2}},\quad q\to 1.
\end{equation}
The order of divergency, also in presence of the halo, is equal to that given 
in H90, and assuming $\mu=0$ or $\beta=1$ the correct asymptotic formula for 
the H model is recovered.\footnote{All the higher-order terms in equations 
(21)-(22) can written explicitly, but their expressions become increasingly 
complicated. Note that equation (21) of H90 is incorrect, as can be shown 
expanding equation (17) there. Note also that the parameter $q$ used in H90 is
the square root of the parameter used in this paper.}

Consistently with the previous discussion the asymptotic formula for $q\to 1$ 
in the case of a central BH {\it cannot} be obtained as the limiting case of 
the previous formula when $\beta=0$, due to a change in the order of the 
singularity in equation (B2). The correct treatment in this case gives:   
\begin{equation}
\ftil (q)\sim {1\over 2\pi^2\sqrt{2\mu q}}{\sqrt{1-q}\over\mu +(1-q)^2},
\quad q\to 1,
\end{equation}
and the convergence to 0 of the DF is proved. Finally, note how the presence of
anisotropy does not affect the behavior of the DF for high relative energies,
as dictated by equation (12), where it can be easily seen that the divergence 
of the modified density near the model center -- the leading term in the 
asymptotic expansion of the DF -- is only due to its isotropic part. The 
asymptotic formulae (21)-(22) have been checked numerically for many choices of
$(\mu,\beta)$ by direct comparison with the DF, and the agreement is excellent.

\placefigure{fig3}

\subsection{Velocity Sections of the DF}

The representation of the DF as function of the integral of motion $Q$ is not 
easily interpreted when orbital anisotropy is present. More intuitive are 
the {\it velocity sections} of $f$, i.e., for some fixed $r$ the 
distributions of $\vrad$ and $\vtan$, the radial and tangential velocity 
components, respectively. In Fig. 3 the sections $f(r,\vrad,0)$ and 
$f(r,0,\vtan)$ for the H model with $\sa=1$ are shown. Obviously, for global 
isotropy the two 
distributions are equal, and for anisotropic systems they become more and more
similar moving from radii greater than $\ra$ to smaller radii, according to the
radial trend of anisotropy implied by the OM parameterization. As expected, for
a finite value of $\ra$, moving outward the tangential orbits become more and 
more de-populated, and this is compensated by an increase in the number of 
high velocity radial orbits. The presence of a massive diffuse halo does not 
alter qualitatively the distributions, that maintain the same aspect. The only
peculiar characteristic appears again in the case of the 
central BH, when the maximum of the distribution is placed off-center: this 
fact reflects the central behavior of the DF described previously. As a final 
and general remark on these sections, one can note how they depart appreciably
from a Maxwellian distribution.

\subsection{The Differential Energy Distribution of HH Models}

In the case of global isotropy the differential energy distribution $dM/d\eps$
for each component of the HH models can be derived analytically as a function 
of
the parameter $l$. As shown by BT87 (p.243), $dM/d\eps=f(\eps)g(\eps)$, where 
\begin{equation}
g(\eps)=16\sqrt{2}\pi^2\gn\Gtilpm (l)
\end{equation}
is the density of states, and $\gn=\rc^3\psin^{1/2}$. The explicit form for 
$\Gtilpm (l)$ is given in Appendix C, and the sign $\pm$ correspond to the 
case $b>0$ and $-1\leq b <0$ respectively. Again, a first check of the derived
formula can be obtained evaluating analytically its limit for $\mu=0$, and 
recovering the $g(\eps)$ as given by H90; as in the case of the DF, the proof 
is cumbersome, and not shown here. When also a halo is present, the check is 
obtained by direct comparison with the density of states derived by numerical 
integration of equation (C2). Over the whole energy range the curves are 
indistinguishable. $dM/d\eps$ is plotted in Fig. 4: when the halo is more 
extended than the reference component ($\beta >1$, dotted lines), $dM/d\eps$ is
not steadily increasing in the outer part of the system anymore. 

\placefigure{fig4}

\subsection{The DF for Halo Dominated HH Models}

In the case of a halo dominated model, i.e., when the self-gravity of the 
reference component is negligible, the DF and $dM/d\eps$ can be expanded for 
$\mu\to\infty$, and the resulting expressions are combinations of elementary 
functions of $l$. For brevity, and due to its major importance, the explicit 
form of the functions entering the DF are given only.

\subsubsection{The Case $\beta >1$}

In this case, $\Ftilpis$ and $\Ftilpan$ are given by equations (B8)-(B9) with:
\begin{equation}
\Cs(l)={5\over 16}\arccos\left({1\over l}\right)
+\sqrt{\l2-1}\left({5\over 16\l2}+{5\over 24 l^4}+{1\over 6 l^6}\right),
\end{equation}
\begin{equation}
\Cq(l)={3\over 8}\arccos\left({1\over l}\right)
+\sqrt{\l2-1}\left({3\over 8\l2}+{1\over 4 l^4}\right),
\end{equation}
\begin{equation}
\Cd(l)={1\over 2}\arccos\left({1\over l}\right)+{\sqrt{\l2-1}\over 2\l2},
\end{equation}
\begin{equation}
\Vu(\beta,l)=\sqrt{\beta-\l2\over\beta}
\arctan\sqrt{\beta (\l2-1)\over\beta-\l2}.
\end{equation} 
\begin{equation}
\Vd(\beta,l)={(2\beta-\l2)\Vu(\beta,l)\over 2\beta}+
{(\beta-\l2)\sqrt{\l2-1}\over 2\beta(\beta-1)}.
\end{equation}

\subsubsection{The Case $0\leq \beta <1$}

In this case, $\Ftilmis$ and $\Ftilman$ are given by equation (B12)-(B13) with:
\begin{equation}
\Ds(l)={5\over 16}\arccosh\left({1\over l}\right)
+\sqrt{1-\l2}\left({5\over 16\l2}+{5\over 24 l^4}+{1\over 6 l^6}\right),
\end{equation}
\begin{equation}
\Dq(l)={3\over 8}\arccosh\left({1\over l}\right)
+\sqrt{1-\l2}\left({3\over 8\l2}+{1\over 4 l^4}\right),
\end{equation}
\begin{equation}
\Dd(l)={1\over 2}\arccosh\left({1\over l}\right)+{\sqrt{1-\l2}\over 2\l2},
\end{equation}
\begin{equation}
\Xu(\beta,l)=\sqrt{\l2-\beta\over\beta}
\arctan\sqrt{\beta (1-\l2)\over\l2-\beta}.
\end{equation}
\begin{equation}
\Xd(\beta,l)={(2\beta-\l2)\Xu(\beta,l)\over 2\beta}+
{(\beta-\l2)\sqrt{1-\l2}\over 2\beta(\beta-1)}.
\end{equation}
In the particular case of a dominating central mass, the DF is obtained setting
$\beta=0$ in equations (B12)-(B13), eliminating the term containing $\Xd$, and
using equations (29)-(32) with $\lim_{\beta\to 0}\Xu(\beta,l)=\sqrt{1-\l2}$.
 
\section{Consistency of HH Models}

We can now explore the parameter space of HH models, studying numerically the 
formulae derived in \S 4. The simplest way to summarize the results is to 
express the consistency limitations in terms of the (normalized) anisotropy 
radius of the reference component. This is particularly indicated because: 
1) we know from the introductory discussion ($\S$ 3.2) that the globally 
isotropic component of the HH model is consistent whatever the halo component 
is (i.e., $\fis>0$), and 2) in the case of the OM anisotropy, the anisotropy 
radius can be isolated in the DF. So, in the parameter space $(\sa,\mu,\beta)$
it is easy to determine the critical value $\sac (\mu,\beta)$ defined as the 
anisotropy radius such that for $\sa <\sac$ there exists at least one permitted
value of the potential for which $f<0$. Imposing the positivity of $f$ over all
the domain $0<q<1$, from equation (19) one obtains:
\begin{equation}
\sac^2(\mu,\beta)={\rm sup}\left[-{f_{\rm a}(q)\over 
f_{\rm i}(q)}\right ]_{q\in ]0,1[}.
\end{equation}

In Fig. 1 different curves $\sac^2$ are plotted, for a varying halo mass, and 
for fixed values of $\beta$. As expected no model permits a smaller value of 
the anisotropy radius than that derived from the necessary condition. The 
intermediate solid line is the value of the exact limit on the anisotropy 
radius for the H model, $\sa\simeq 0.202$. The upper solid curve is the plot of
the WSC in the case of a central BH of normalized mass $\mu$. As described in 
\S 3.2 this curve approaches asymptotically the value $1/\sqrt{2}$.

The dashed and dotted lines represent the lower bound on the anisotropy radius
for various $\beta$. Clearly all the families of curves for $\mu\to 0$ converge
to the value required by the H model. The first result is that the critical 
anisotropy radius for each model is not strongly dependent on the halo mass, 
and it is always contained between the value given by the necessary condition 
and that obtained from the SSC applied to the H+BH model. The second result is
that all models with $0\leq \beta <1$ (short-dashed lines) have a critical 
anisotropy radius higher than the H model: as expected a model with a 
concentrated halo cannot sustain also too much anisotropy. The long-dashed line
represents the case of a central BH: anisotropy radii higher than the values 
represented by this curve can be assumed independently of the halo structure 
and mass. The third result is that all models in which the halo is more 
extended than the reference component can have a slightly smaller anisotropy 
radius than the H model, but in this case the effect is much less stronger than
for a $\beta <1$ halo: as already found in CP92, the most diffuse component of
a multi-component system is also the most "delicate" concerning the 
consistency. Finally, note how all the curves become flat asymptotically. Their
limiting value cannot be obtained directly using equation (19) for very high 
$\mu$ values, due to the numerical precision loss of the intervening functions.
On the contrary a very accurate analysis can be performed using the asymptotic
functions given in \S 4.3, and the result is plotted in Fig. 5. 

\placefigure{fig5}

\subsection{Stability of HH Models}

As important as the discussion on the consistency of HH models is the location
of the region of {\it stability} for such systems. A complete stability 
analysis is beyond the task of this work, requiring N-body simulations or 
normal mode analysis, but some interesting conclusions can be equally derived,
at least for the globally isotropic case. In fact in this case powerful
theorems are at our disposition: for stability against both radial and 
nonradial perturbations it is sufficient (but not necessary!) that the system 
DF (in our case the sum of the DF of the halo and of the reference component) 
is an increasing function of the relative binding energy $\eps$ (see, e.g., 
BT87, p.296-307, \cite{fripoly84}, p.152-163). 

In the performed numerical exploration of the parameter space $(\mu,\beta)$ all
the DFs {\it with} $\beta>0$ are monotonically increasing functions of the 
parameter $q$, as shown for some particular cases in Fig. 2. Changing the 
parameter $q$ of each component to $\eps$ using equation (17), and taking the 
derivative shows that the {\it globally isotropic HH models are stable}. 
Unfortunately the sufficient condition above cannot be applied to the globally
isotropic case with a central BH: the condition $df/d\eps >0$ is not verified,
due to the convergence of the DF to 0, and so only numerical investigations can
answer this interesting problem.  

For anisotropic systems the situation is more complex, for the lack of general 
theorems. In any case some hint can be obtained by the empirical requirement 
that $\xi\equiv 2\Krad/\Ktan\lsim\csic=1.7\pm 0.25$ (\cite{fripoly84}, p.235) 
in order to avoid the radial orbit instability, where $\Krad$ and $\Ktan$ are 
the total radial and tangential kinetic energies of the system, and the exact 
value of $\csic$ is model dependent. A second complication with the previous 
criterion arises form the fact that a generalization to multi-component systems
is not obvious, as shown by Stiavelli and Sparke (1991). In the same paper the
authors show with the aid of N-body simulations that the presence of a halo 
does not change very much the situation with respect to the one-component 
model. For HH models with OM anisotropy the evaluation of $\Krad$ and $\Ktan$ 
can be done analytically, but the resulting formulae for $\beta >0$ are 
complicated. For these reasons I show here only the trend of the parameter 
$\xi(\sa,\mu)$ for the particular case of the H+BH system: the simpler case of
an H model is obtained for $\mu=0$. From the virial theorem $\Krad +\Ktan=
(|U|+|W|)/2$, where $U$ is the gravitational energy of the H model, $W$ is its
interaction energy with the BH, and so $\xi=2/[(|U|+|W|)/2\Krad -1]$. Choosing
as the energy normalization $\un=M\psin$, it results $U=-2\pi\int\rho\Psi r^2dr
=-\un/6$, $W=4\pi\int r^2\rho (d\psih/dr)dr=-\un\mu$, and $\Krad=2\pi\int\rho
\sigma^2_{\rm r}r^2dr=\un(\tilde\KradH+\mu\tilde\KradBH)$, with
\begin{eqnarray}
\tilde\KradH & = & {1+8\sad+23\saq+12\sas\over 12(1+\sad)^2}+
                   {\pi\sa (-1+16\sad+15\saq+6\sas)\over 24(1+\sad)^3}+
                   \nonumber\\
             &   & \ln(\sa)\sat
                   \left[{\sa(7+8\sad+3\saq)\over 3(1+\sad)^3}+{\pi\over 2}+
                   \arctan (\sa)\right]-\nonumber\\
             &   & {\sat [\pi\ln(1+\sad)+\sa\Phi(-\sad,2,1/2)]\over 4},
\end{eqnarray}
and
\begin{eqnarray}
\tilde\KradBH &=& {12\sad+1\over 2}+{\pi\sa(2+3\sad)\over 2(1+\sad)}+
                  \ln(\sa)\sa 
                   \left\{{\sa (5+6\sad)\over 1+\sad}+(1+6\sad)\left(
                         {\pi\over 2}+\arctan (\sa)\right)\right\}-\nonumber\\
              & &{\sa(1+6\sad)[\pi\ln(1+\sad)+\sa\Phi(-\sad,2,1/2)]\over 4}.
\end{eqnarray}
I have used for $\sigma_{\rm r}^2(r)$ required in the integration the 
expression given in Ciotti et al. (1996) [eqs. (A1)-(A3)], and the function 
$\Phi$ is the Lerch transcendent (see, e.g., \cite{emot53} vol.1, p.27). In 
Fig. 6 the function $\xi$ is plotted for various $\mu$ values, and the 
asymptotic flattening to unity for increasing isotropy is evident. The first 
comment is that the stability criterion requires, also in the conservative 
hypothesis of a very high value of $\csic$ (i.e. $\simeq 2.5$), minimum 
anisotropy radii appreciably larger than those obtained form the consistency 
analysis. For example, the minimum anisotropy radius permitted for an H model 
is $\simeq 0.93$, much higher than the value required by the simple 
consistency, $\simeq 0.2$. So, it is likely that the more radially anisotropic
HH models with positive DF are with any probability radial orbit instability 
prone. Another important comment concerns the relation between the relative 
distribution of the halo and of the reference component. As can be seen from 
Fig. 6, haloes more extended than the reference component make the model more 
unstable; on the contrary, more stable systems are obtained for more 
concentrated haloes. This is easily explained looking at the radial trend of 
the velocity dispersion as a function of the anisotropy radius. For extended 
haloes, the velocity dispersion is mainly increased in the outer parts of the 
model, where also the orbits are strongly radial, and this correspondingly 
increases the $\xi$ value; the opposite happens for very concentrated haloes. 
Particularly explicit are the two lines in Fig. 6 referring to a central BH: in
this case the velocity dispersion is essentially increased only at the model 
center, where isotropy is nearly realized, and the stability indicator remains
low also for small values of $\ra$. But in this case probably the global 
indicator $\xi$ looses its meaning, due to a very strong {\it decoupling} 
between the central and the outer parts of the model.  

\placefigure{fig6}

\section{Conclusions}

In this paper an extensive analytical investigation of two-component spherical
galaxy (or cluster) models, made of the sum of two Hernquist density 
distributions with different physical scales, is carried out. A variable amount
of orbital anisotropy is also allowed in both components. These models,
characterized by a power-law density profile in their central regions -- both 
in the visible and in the dark matter distribution -- reproduce the main 
properties of early-type galaxies as revealed by Hubble Space Telescope 
observations, and also of the dark matter distribution as obtained in recent 
N-body simulations. The main results presented in this paper can be summarized
as follows:

\begin{enumerate}
\item The analytical expression for the DF of HH models with general OM 
anisotropy is presented and discussed, even for the particular case of an
Hernquist model with a central BH. The special case of a dominant dark halo is
also discussed, and it is shown that under this assumption the DF can be 
asymptotically expressed using just elementary functions. In the case of global
isotropy the analytical expression for the differential energy distribution of 
both components is obtained. Some velocity sections of the DF are shown and 
discussed.

\item The necessary and sufficient conditions that the model parameters must 
satisfy in order to correspond to a consistent system (i.e., a system for which
each physically distinct component has a positive DF) are analytically derived
using the method introduced in CP92. It is proved that globally isotropic HH 
models are consistent for any mass ratio and core radii ratio, even in the
special case in which the "halo" reduces to a BH. In the case of a central BH 
and of variable anisotropy for the host system, the analytical expression for 
a minimum anisotropy radius as a function of the BH mass is given. 

\item The region in the parameter space in which HH models are consistent is 
subsequently explored using the DF. The main result is that the presence of a 
massive halo does not affect significantly the maximum anisotropy that can be 
sustained by a consistent model. It is shown that the presence of a halo with 
a core radius larger than that of the reference component allows a slightly 
higher degree of anisotropy with respect to the one-component Hernquist model.
On the contrary, a halo with a smaller core radius imposes a larger value for 
the minimum anisotropy radius than that proper of the H model. The most 
restrictive case is that of a central BH. In any case, for a given core radius
of the halo there is a lower limit to the minimum anisotropy radius that 
approaches an asymptotic value for a dominant halo mass. 

\item Finally, it is proved that isotropic HH models are stable, except for the
case of a central BH, when no conclusions can be drawn. For anisotropic models
the stability parameter against radial orbit instability is briefly discussed,
and it is shown that with high probability the most anisotropic 
{\it consistent} HH models are unstable. 
\end{enumerate}

\acknowledgments

I would like to thank Giuseppe Bertin and Silvia Pellegrini for helpful 
comments and discussions. This work has been partially supported by the 
Italian Ministry of Research (MURST).

\appendix
\section{Consistency Requirements}

In this Appendix I apply the CP92 method to the HH models, as discussed 
in $\S$ 3. The necessary condition for the H model imposes through 
equation (4) a limitation on the anisotropy radius:
\begin{equation}
\sa^2\geq {(3\psitil-2)(1-\psitil)^2\over (4-3\psitil)\psitil^2},
\quad  0\leq\psitil\leq 1.
\end{equation}
The requirement is that $\sa^2$ is larger than or equal to the maximum of the
function on the r.h.s.. This is reached at
\begin{equation}
\psitilm={5\over 4}-{\sqrt{33}\over 12}\simeq 0.77,
\end{equation}
that, after back substitution in equation (A1), gives equation (13). 

The SSC applied to the H model gives the following inequality:
\begin{equation}
\sa^2\geq {3(5\psitil-2)(1-\psitil)^3
\over (15\psitil^2-39\psitil+28)\psitil^2},\quad 0\leq\psitil\leq 1.
\end{equation}
After differentiation, discarding the two complex conjugates roots of the 
resulting cubic equation, and defining 
\begin{equation}
\phi_0=\left({1336\over 3645}+{\sqrt{35887965}\over 16200}\right)^{1/3},
\end{equation}
the maximum of the r.h.s. of equation (A3) is reached at
\begin{equation}
\psitilm={23\over 18}+{217\over 1620\phi_0}-\phi_0\simeq 0.52,
\end{equation}
that after back substitution in equation (A3) gives equation (14).

The WSC applied to the H+BH model is obtained from equation (6) with 
$b=-1$. The inequality to be verified for a given $\mu$ is:
\begin{equation}
\sa^2\geq-{3\psitil^4-10\psitil^3+6(2+\mu)\psitil^2-6(1+\mu)\psitil+(1+\mu)
\over (3\psitil^2-8\psitil+6+6\mu)\psitil^2},\quad 0\leq\psitil\leq 1. 
\end{equation}
Taking the derivative of the r.h.s. we are left 
with the discussion of a rational function whose denominator is strictly 
positive, and the numerator is a polynomial of fifth degree, that fortunately 
can be factorized in a term of second degree strictly positive, and the cubic:
\begin{equation}
\cmu (\psitil)=-\psitil^3+4\psitil^2-6(1+\mu)\psitil+2(1+\mu).
\end{equation}
Observing that $\cmu(0)=2(1+\mu)>0$ and $\cmu(1)=-(1+4\mu)$, there is at least 
one solution of $\cmu (\psitil)=0$ between 0 and 1. This solution is also the 
only one, because the determinant of the cubic $\cmu$
\begin{equation}
\Delta_{\mu}={4(1+\mu)(216\mu^2+99\mu+11)\over 27},
\end{equation}
is positive $\forall\mu\geq 0$: from the theory of algebraic equations, 
$\cmu (\psitil)=0$ admits one and only one real solution. Finally, considering
the sign of $\cmu$ we proved that the solution $\psitilm$ corresponds to a 
maximum. Defining 
\begin{equation}
\phi_{\mu}=\left[{\sqrt{3(1+\mu)(216\mu^2+99\mu+11)}\over 9}-3\mu-
{17\over 27}\right]^{1/3},
\end{equation}
one obtains
\begin{equation}
\psitilm(\mu)={4\over 3}-{2(1+9\mu)\over 9\phi_{\mu}}+\phi_{\mu},
\end{equation}
that after substitution in equation (A6) gives equation (15). 
When $\mu=0$, the value at which the maximum is reached can be calculated 
directly from equations (A9)-(A10), obtaining:
\begin{equation}
\phi_0=\left[{\sqrt{33}\over 9}-{17\over 27}\right]^{1/3},
\end{equation}
and
\begin{equation}
\psitilm={4\over 3}-{2\over 9\phi_0}+\phi_0\simeq 0.456,
\end{equation}
that after substitution in eq. (A6) with $\mu=0$ gives eq. (16).

The application of the WSC to the globally isotropic HH models is more 
complicated. After having computed the derivatives in equation (6), we have to 
investigate the positivity of a rational expression, whose denominator is
strictly positive $\forall b\geq -1$, $b\neq 0$, $\forall\mu\geq 0$ and 
$0\leq\psitil\leq 1$. The numerator of this function factorizes in a strictly 
positive function and in the polynomial: 
\begin{equation}
\mu c_{b}(\psitil)+(3\psitil^2-8\psitil+6)(1+b\psitil)^3\geq 0,
\end{equation}
where
\begin{equation}
c_{b}(\psitil)=6b\psitil^3+3(1-5b)\psitil^2+2(5b-4)\psitil+6.
\end{equation}

It is trivial to show that the second addend of equation (A13) is strictly 
positive. I prove now that also $c_{b}$ is positive for $b\geq -1$ and for 
$0\leq\psitil\leq 1$: so the WSC is satisfied for any choice of $(\mu;\beta)$ 
and the globally isotropic HH models are consistent. First of all, 
$c_{b}(0)=6$ and $c_{b}(1)=b+1=\beta\geq 0$, and so after excluding the 
presence of roots of $c_{b}$ in the interval [0,1], the proof is obtained. This
can be accomplished using the classical Sturm method, i.e., counting the 
differences in the number of variations between 0 and 1 in the Sturm sequence 
$S_b$ associated to $c_b(\psitil)$ and discussing it as a function of the 
parameter $b$. But a faster proof can be obtained as follows. Since $c_{b}$ 
does not change sign between 0 and 1, only an {\it even} number of roots can be
contained in the interval. If we show that equation (A14) admits only one real
solution then the proof is obtained. The discriminant of $c_{b}$
\begin{equation}
\Delta_b={(125b^2+50b+6)(b+1)^2\over 2916b^4},
\end{equation}
is strictly positive, and so the cubic equation admits one and only one real 
solution, that is necessarily placed outside the interval [0,1], and 
$c_b(\psitil)\geq 0$ for $0\leq\psitil\leq 1$.

\section{The DF for HH Models}

In this Appendix the main analytical steps required for the determination of
the DF are described. As discussed in \S 4, the first step is to change the 
integration variable from the total potential to the potential of the reference
component. After a normalization, equation (2) becomes: 
\begin{equation}
f(Q)={\fn\over\sqrt{8}\pi^2}{d\Ftil[q(\Qtil)]\over d\Qtil},
\end{equation}
where the relation between $q$ and $\Qtil$ is given by equation (17), and
\begin{equation}
\Ftil(q)=\int_0^q{d\tilde\varrho\over d\psitil}
{d\psitil\over\sqrt{\psitilt (q)-\psitilt (\psitil)}}.
\end{equation}
From equations (11) and (17)
\begin{equation}
\psitilt (q)-\psitilt(\psitil)={(q-\psitil)(\d2+1+b\psitil)\over 1+b\psitil},
\quad\d2\equiv {\mu\over\l2}.
\end{equation}
It is shown in the next subsections that $\Ftil$ is actually a function of $q$
only through $l$ defined by equation (18). Unfortunately its form changes 
depending on the sign of $b=\beta -1$, and it is convenient to separate the 
discussion of the cases $\beta >1$ and $0\leq\beta <1.$

\subsection{The Case $\beta>1$}

This is the simpler case, for which $b>0$ and $1\leq\l2\leq\beta$. After the 
change of variable 
\begin{equation}
t=\sqrt{1+b\psitil},
\end{equation}
equation (19) becomes:
\begin{equation}
\Ftilp(l)={2\over b^{5/2}}\int_1^l
\left[\rapis (t)+{\rapan (t)\over\sa^2}\right]
{dt\over\sqrt{(\d2+t^2)(\l2-t^2)}}.
\end{equation}  
After a conversion in simple fractions it results:
\begin{equation}
\rapan (t)=-3t^6+2(2+\beta)t^4-(1+2\beta)t^2,
\end{equation}
and 
\begin{equation}
\rapis (t)=-3t^6-2(\beta-4)t^4-(\beta^2-4\beta+6)t^2-
{(\beta-1)^4\over\beta-t^2}+{\beta (\beta-1)^4\over (\beta-t^2)^2}.
\end{equation}

Using the nomenclature of Byrd and Friedman (\cite{bf71}, hereafter BF71) we 
have:
\begin{equation}
\Ftilpis(l)={2g\over b^{5/2}}
\left [-3l^6 \Cs -2(\beta-4)l^4 \Cq  -(\beta^2-4\beta+6)\l2\Cd -
{(\beta-1)^4\over \beta -\l2}\Vu + {\beta (\beta-1)^4
\over (\beta -\l2)^2}\Vd \right],
\end{equation}
and for the anisotropic part,
\begin{equation}
\Ftilpan(l)={2g\over b^{5/2}}
[-3l^6 \Cs +2(\beta+2)l^4 \Cq  -(1+2\beta)\l2\Cd],
\end{equation}
where the functions $\Cm (\phi,k)$ and $\Vm(\phi,\al2,k)$ are given in 
Appendix D, and
\begin{equation}
g={1\over\sqrt {\d2+\l2}},\quad\k2={\l2\over \d2+\l2},
\quad\phi=\arccos\left({1\over l}\right),
\quad\al2 ={\l2\over \l2 -\beta},
\end{equation}
(BF71, p.48).

\subsection{The Case $0\leq\beta <1$}

In this case $\beta\leq\l2\leq 1$. After the change of variable given in 
equation (B4), equation (19) becomes:
\begin{equation}
\Ftilm(l)={2\over \vab^{5/2}}\int_l^1
\left[\rapis (t)+{\rapan (t)\over\sa^2}\right]
{dt\over\sqrt{(\d2+t^2)(t^2-\l2)}}. 
\end{equation}  
For the isotropic part:
\begin{equation}
\Ftilmis(l)={2g\over\vab^{5/2}}
\left [-3l^6 \Ds -2(\beta-4)l^4 \Dq  -(\beta^2-4\beta+6)\l2\Dd -
{(\beta-1)^4\over \beta -\l2}\Xu + {\beta (\beta-1)^4
\over (\beta -\l2)^2}\Xd \right],
\end{equation}
and for the anisotropic part
\begin{equation}
\Ftilman(l)={2g\over\vab^{5/2}}
[-3l^6 \Ds +2(\beta+2)l^4 \Dq  -(1+2\beta)\l2\Dd],
\end{equation}
where the functions $\Dm(\phi,k)$ and $\Xm(\phi,\al2,k)$ are given in Appendix
D, and 
\begin{equation}
g={1\over\sqrt {\d2+\l2}},\quad\k2={\d2\over \d2+\l2},
\quad\phi=\arccos (l),\quad\al2 ={\beta\over \beta-\l2},
\end{equation}
(BF71, p.45).

The particular case of a central BH can be derived from the previous formulae.
In this case $b=-1$ and so $0\leq\l2\leq 1$. The coefficients of the special 
functions are obtained from equations (B12)-(B13) for $\beta =0$, and their 
arguments are still given by equation (B14). The only problem is in the 
isotropic part: the coefficient of $\Xd$ is zero and 
$\lim_ {\al2\to 0}\Xu(\phi,\al2,k)=\Cd(\phi,k)$. The anisotropic part is 
obtained by direct substitution of $\beta=0$ in equation (B13).

\section{The Density of States}

The formula (4.157b) of BT87, after normalization of radii and energy to the 
core radius and central potential of the reference component can be written as:
\begin{equation}
\Gtil(\Etil)=\int_0^{s_{\rm M}(\Etil)}s^2\sqrt{\psitilt (s)-\Etil}ds,
\end{equation}
with $\psitilt(s_{\rm M})=\Etil$. Changing the variable of integration from the
radius $s$ to the relative potential of the reference component, the result
is: 
\begin{equation}
\Gtil(q)=\int_q^1\sqrt{\psitilt (\psitil)-\psitilt (q)}
{(1-\psitil)^2\over\psitil^4}d\psitil,
\end{equation}
where the relation between $q$ and $\Etil$ is given by equation (17). Again
as for the DF it is convenient to discuss separately the two cases $\beta>1$ 
and $0\leq\beta <1$.

\subsection{The Case $\beta >1$}

Changing the variable of integration as in equation (B4), from 
equation (C2)  
\begin{equation}
\Gtilp (l)=2b^{1/2}\int_l^{\sqrt{\beta}}{\gap (t)dt\over
\sqrt{(\delta^2+t^2)(t^2-l^2)}},
\end{equation} 
where
\begin{eqnarray}
\gap (t) & = & 1+{2\beta-4+\l2-\d2\over 1-t^2}+
               {(\d2-\l2)(3-2\beta)+\beta(\beta-6)+6-\mu\over (1-t^2)^2}+
               \nonumber\\
         &   & {(1-\beta)[(\beta-3)(\d2-\l2)+2(\beta+\mu-2)]\over (1-t^2)^3}+
               {(1-\l2)(1-\beta)^2(1+\d2)\over (1-t^2)^4}.
\end{eqnarray}
After integration, 
\begin{eqnarray}
{\Gtilp (l)\over 2b^{1/2}g} & = & \Xz+{2\beta-4+\l2-\d2\over 1-\l2}\Xu+
         {(\d2-\l2)(3-2\beta)+\beta(\beta-6)+6-\mu\over (1-\l2)^2}\Xd+
         \nonumber\\
                            &   & {(1-\beta)[(\beta-3)(\d2-\l2)+2(\beta+\mu-2)]
         \over (1-\l2)^3}\Xt+{(1-\beta)^2(1+\d2)\over (1-\l2)^3}\Xq,
\end{eqnarray}
where the functions $\Xm$ are given in Appendix D, and their arguments are 
\begin{equation}
g={1\over\sqrt {\d2+\l2}},\quad\k2={\d2\over \d2+\l2},
\quad\phi=\arccos \left({l\over\sqrt{\beta}}\right),
\quad\al2 ={1\over 1-\l2}.
\end{equation}

\subsection{The Case $0\leq\beta <1$}

In this case, 
\begin{equation}
\Gtilm (l)=-2\vab^{1/2}\int_{\sqrt{\beta}}^l{\gap (t)dt\over
\sqrt{(\delta^2+t^2)(l^2-t^2)}},
\end{equation} 
where $\gap (t)$ is given in eq. (C4). 
After integration, 
\begin{eqnarray}
{\Gtilm (l)\over -2\vab^{1/2}g} & = &\Vz+{2\beta-4+\l2-\d2\over 1-\l2}\Vu+ 
     {(\d2-\l2)(3-2\beta)+\beta(\beta-6)+6-\mu\over (1-\l2)^2}\Vd+\nonumber\\
                                &   &{(1-\beta)[(\beta-3)(\d2-\l2)+
     2(\beta+\mu-2)]\over (1-\l2)^3}\Vt+{(1-\beta)^2(1+\d2)\over (1-\l2)^3}\Vq,
\end{eqnarray}
where the functions $\Vm$ are given in Appendix D, and their arguments are
\begin{equation}
g={1\over\sqrt {\d2+\l2}},\quad\k2={\l2\over \d2+\l2},
\quad\phi=\arccos\left({\sqrt{\beta}\over l}\right),
\quad\al2 ={\l2\over \l2 -1}.
\end{equation}
The case of the central BH is obtained from the previous two equations with 
the substitution $\beta =0$.

\section{Elliptic and Jacobian Functions}

Here I shortly summarize the notation for the special functions used, and their
mutual relations, following BF71. The zero-th order functions are:
\begin{equation}
\Cz (\phi,k)=\Dz (\phi,k)=\Vz (\phi,k)=F(\phi,k).
\end{equation}
The m-th order functions are given by:
\begin{equation}
\Cd (\phi,k)={E(\phi,k)-(1-\k2)\Cz\over\k2},
\end{equation}
\begin{equation}
C_{\rm 2m+2}(\phi,k)={2m(2\k2-1)C_{\rm 2m}+(2m-1)(1-\k2)C_{\rm 2m-2}+
\sn(u)\dn(u)\cn^{2m-1}(u)\over (2m+1)\k2},
\end{equation}
(BF71, eq. 213.06-312.05).
\begin{equation}
\Vu (\phi,\al2,k)=\Pi (\phi,\al2,k),
\end{equation}
\begin{eqnarray}
\Vd (\phi,\al2,k) & = & {\al2 E(\phi,k)+(\k2-\al2)\Vz+(2\al2\k2+2\al2-
    \alpha^4-3\k2)\Vu\over 2(\al2-1)(\k2-\al2)}-
    \nonumber\\   &   &
    {\alpha^4\sn(u)\cn(u)\dn(u)\over 2(\al2-1)(\k2-\al2)[1-\al2\sn^2(u)]},
\end{eqnarray}
\begin{eqnarray}
V_{\rm m+3}(\phi,\al2,k) & = & 
        {(2m+1)\k2\Vm+2(m+1)(\al2\k2+\al2-3\k2)V_{\rm m+1}\over
        2(m+2)(1-\al2)(\k2-\al2)}+\nonumber\\ &  & 
        {(2m+3)(\alpha^4 -2\al2\k2-2\al2+3\k2)V_{\rm m+2}\over
        2(m+2)(1-\al2)(\k2-\al2)}+\nonumber\\ &  &
        {\alpha^4\sn(u)\cn(u)\dn(u)\over 2(m+2)(1-\al2)(\k2-\al2)
        [1-\al2\sn^2(u)]^{m+2}},
\end{eqnarray}
(BF71, eq. 213.11-336.03).
\begin{equation}
\Dd (\phi,k)={(1-\k2)\Dz-E(\phi,k)+\tn(u)\dn(u)\over 1-\k2},
\end{equation}
\begin{equation}
D_{\rm 2m+2}(\phi,k)={2m(1-2\k2)D_{\rm 2m} +(2m-1)\k2D_{\rm 2m-2}+
\tn(u)\dn(u)\nc^{2m}(u)\over (2m+1)(1-\k2)},
\end{equation}
(BF71, eq. 211.09-338.05).
\begin{equation}
\Xm (\phi,\al2,k)={1\over\alpha^{2m}}
\sum_{j=0}^m (\al2-1)^j {m!\over j!(m-j)!}V_j,
\end{equation}
(BF71, eq. 211.14-338.04).

In the previous expressions, the elliptic integrals of first, second and 
third kind are expressed as functions of the phase $\phi$ and the modulus $k$:
\begin{equation}
F(\phi,k)=\int_0^{\phi}{d\theta\over\sqrt{1-\k2\sin^2\theta}},
\end{equation}
\begin{equation}
E(\phi,k)=\int_0^{\phi}\sqrt{1-\k2\sin^2\theta}d\theta,
\end{equation}
\begin{equation}
\Pi(\phi,\alpha^2,k)=\int_0^{\phi}{d\theta\over (1-\alpha^2\sin^2\theta)
\sqrt{1-k^2\sin^2\theta}}.
\end{equation}

The relation of the elliptic integrals with the Jacobian functions for given
$(\phi,k)$ are:
\begin{equation}
u=F(\phi,k);\quad\cn(u)=\cos(\phi);\quad\sn(u)=\sin(\phi);
\quad\dn(u)=\sqrt{1-\k2\sn^2(u)},
\end{equation}
and 
\begin{equation}
\quad\tn(u)={\sn(u)\over\cn(u)};\quad\nc(u)={1\over\cn(u)}.
\end{equation}

\clearpage

\figcaption[]{The limits on the anisotropy radius for the consistency of the 
reference component, for various $(\mu,\beta)$ values. The lowest solid line is
the necessary condition. The middle solid line is the minimum anisotropy radius
for the H model. The dotted lines are the limits in the presence of a more 
extended halo, with $\beta=2,5,20$. The short-dashed lines correspond to a 
halo with $\beta=0.5,0.1$, and the long-dashed lines to the case of the central
BH. The dot-dashed and solid line are the SSC and WSC respectively, for the 
case of the central BH.  
\label{fig1}}

\figcaption[]{Upper Panel: the DFs for the reference component of the HH models
in case of global isotropy. The solid line is the DF of the H model. The dotted
lines correspond to the DF in presence of a halo more extended than the 
reference component, with $\beta=5$, and $\mu=5,10,20$. The short-dashed lines
refer to a more concentrated halo, with $\beta=0.2$, and with masses 
$\mu=0.1,1,10$. Finally, the long-dashed lines correspond to a central BH, of 
masses $\mu=0.01,0.1,1.$ Lower Panel: the DFs for the reference component of 
the HH model in the case of radial anisotropy with $\sa=1$. The various curves 
correspond to the same parameters as in the isotropic case. 
\label{fig2}}

\figcaption[]{Radial (solid lines) and tangential (dotted line) normalized
velocity sections for the H model, with $\sa=1$. The numbers near the lines 
give the normalized radius at which the section is shown. 
The dashed line is the 
section at $r/\rc\simeq 0.01$ in the case of a central BH with $\mu=0.1$. 
\label{fig3}}

\figcaption[]{The differential energy distribution for the reference component
in case of global isotropy. The solid line is the $dM/d\eps$ of the H model. 
The other lines correspond to the same parameters described in Fig. 2. 
\label{fig4}}

\figcaption[]{The minimum value for the anisotropy radius in case of a dominant
halo. Note the qualitative difference moving from a less extended halo to a 
more extended halo than the reference component. \label{fig5}}

\figcaption[]{The value of the stability parameter for the reference component
as function of the anisotropy radius. The solid line 
refers to the H model, and the dotted lines to H+BH models: the numbers are the
$\mu$'s values. The other lines are the value of the stability parameter
for various values of $(\mu;\beta)$: circles $(1,0.1)$, triangles $(10,1.5)$, 
and squares $(10;5)$. \label{fig6}}

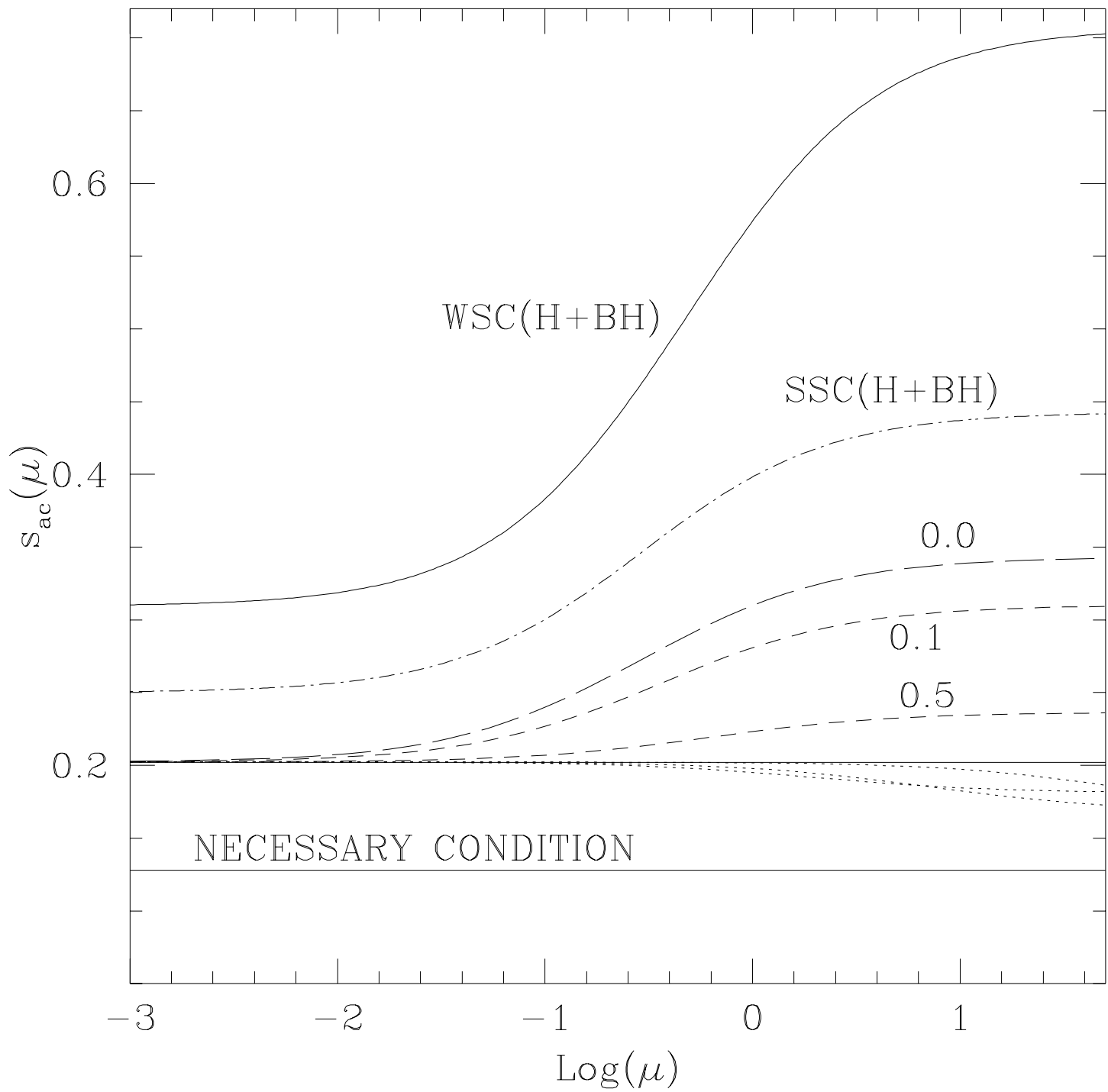
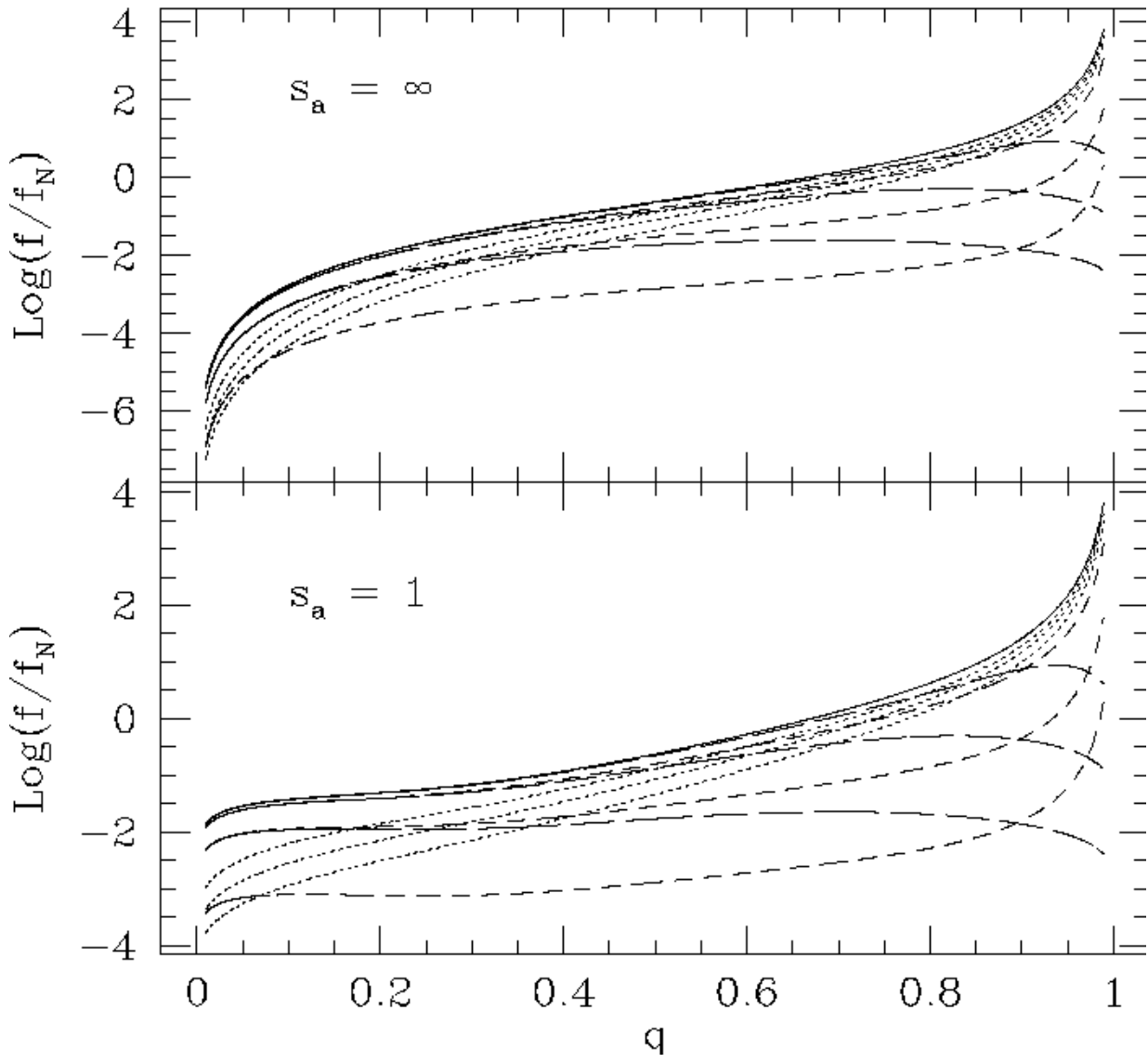
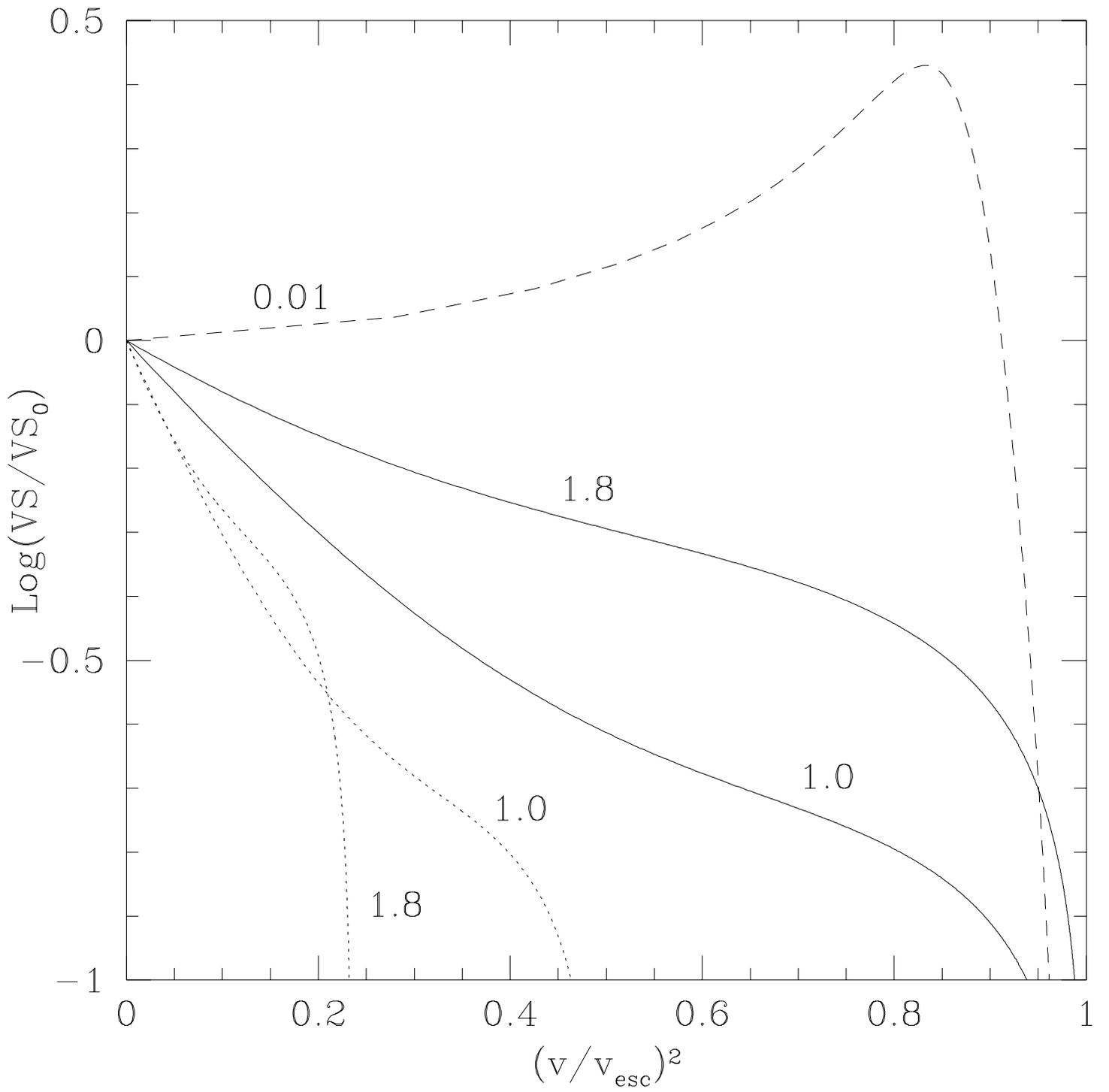
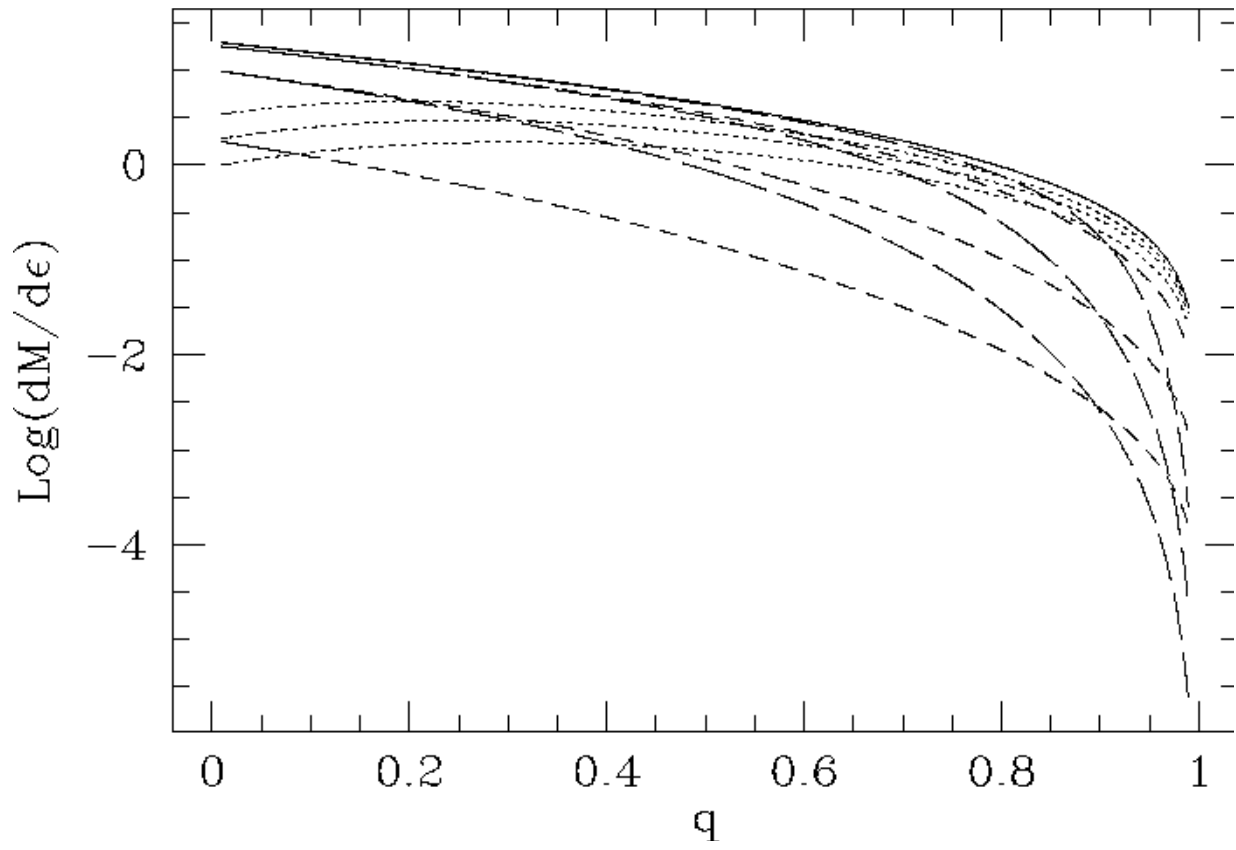
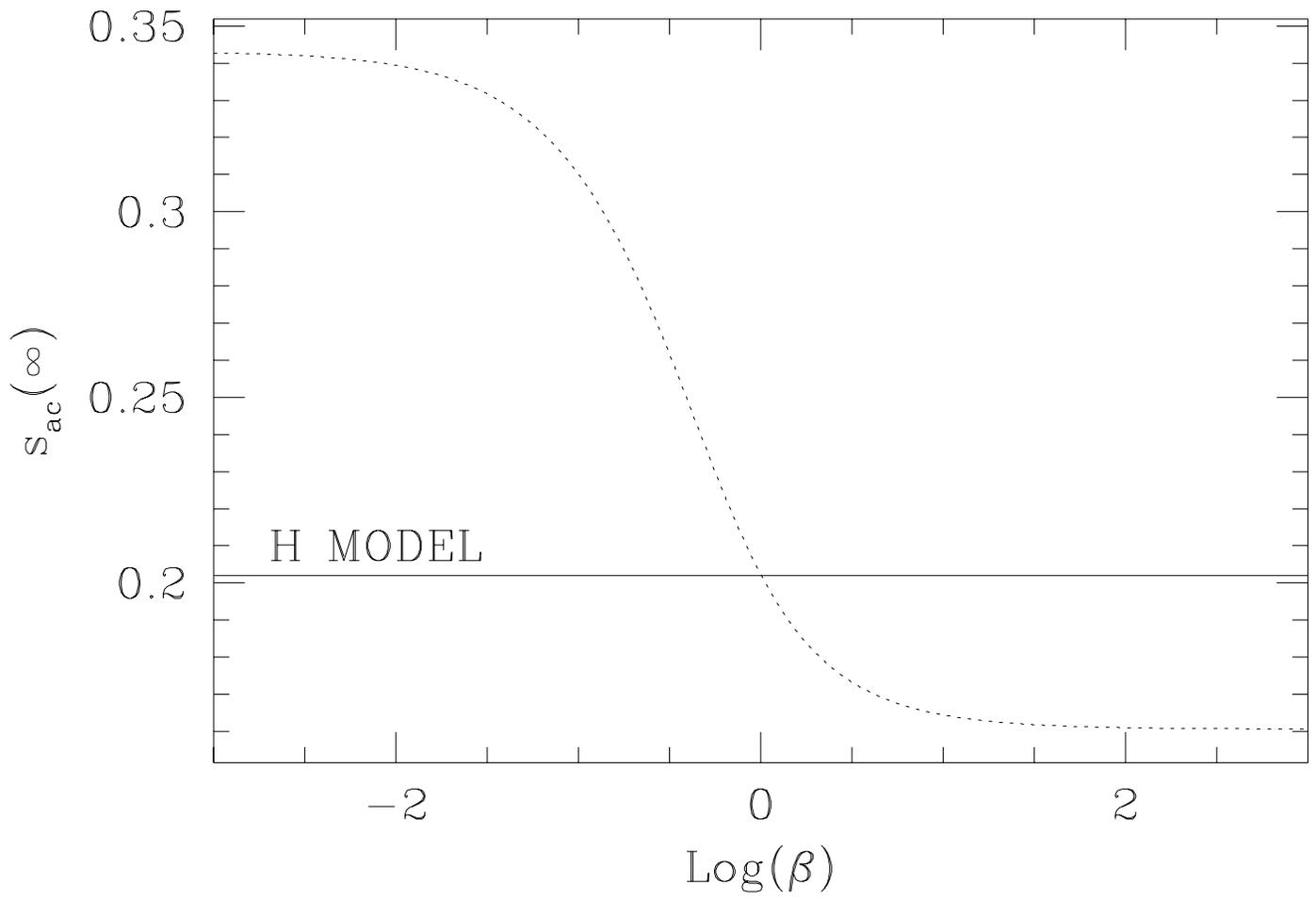
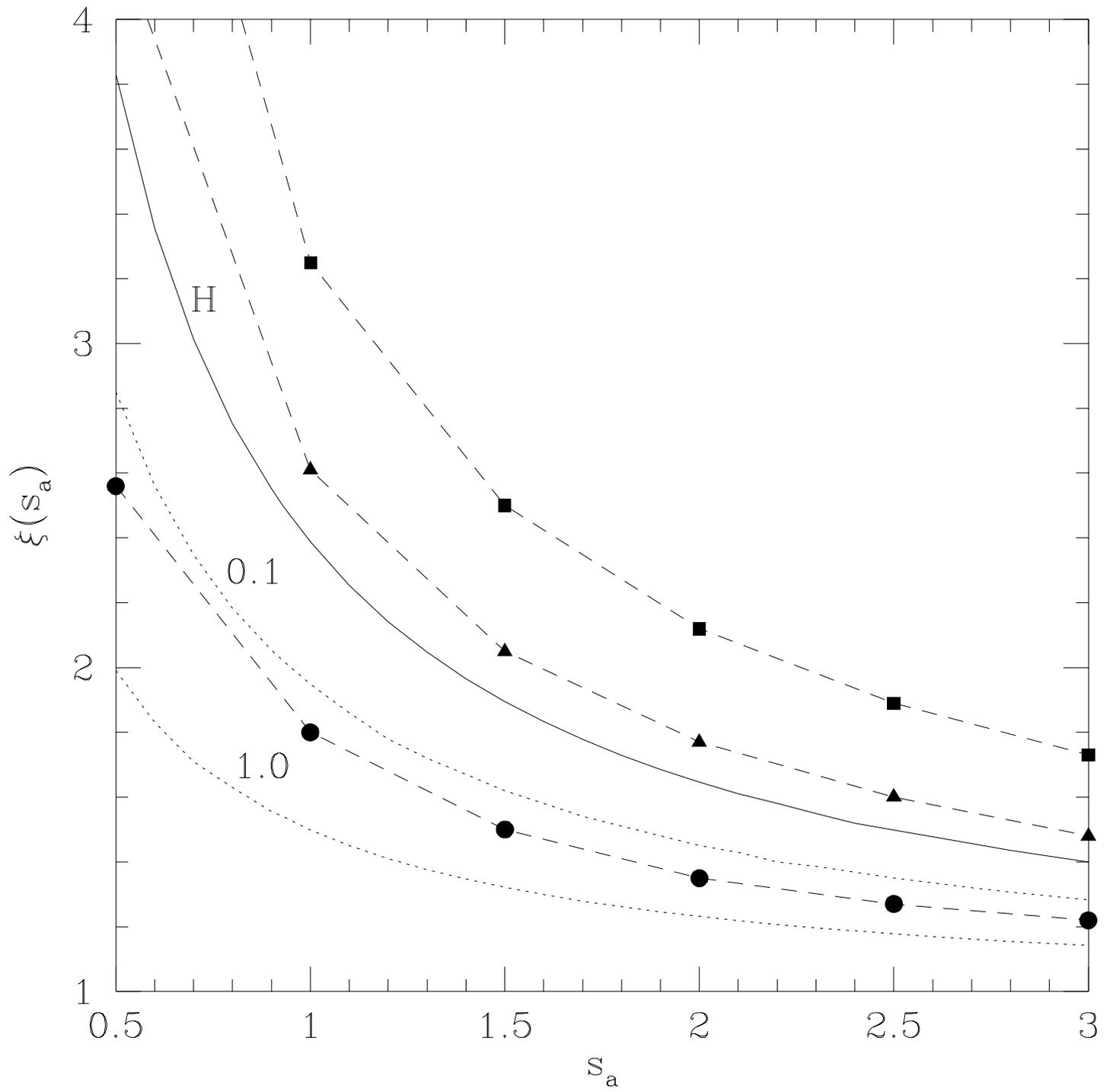
\end{document}